\documentclass{aa}

\usepackage{graphicx}
\usepackage{txfonts}
\usepackage{natbib}
\bibpunct{(}{)}{;}{a}{}{,} 
\usepackage{hyperref}
\hypersetup{colorlinks,
citecolor=blue
}
\begin{document} 

   \title{New red giants in NGC 6791 and NGC 6819 using \emph{Kepler} superstamps}

   \author{A. Covelo-Paz
          \inst{1,2,3}
          \and
          N. Theme{\ss}l\inst{2,3}
          \and
          F. Espinoza-Rojas\inst{2,3}
          \and
          S. Hekker\inst{2,3}
          }

   \institute{Department of Astronomy, University of Geneva, Chemin Pegasi 51, 1290 Versoix, Switzerland\\ 
   \email{alba.covelopaz@unige.ch}
   \and
   Center for Astronomy, Landessternwarte (ZAH/LSW), Heidelberg University, K{\"o}nigstuhl 12, 69117 Heidelberg, Germany
         \and
         Heidelberg Institute for Theoretical Studies (HITS), Schloss-Wolfsbrunnenweg 35, 69118 Heidelberg, Germany\\
}

   \date{Received XX; accepted YY}

 
\abstract{Stars that are members of stellar clusters are assumed to be formed at the same time and place from material with the same initial chemical composition. These additional constraints on the ensemble of cluster stars make these stars suitable as benchmarks.}{We aimed 1) to identify previously unknown red giants in the open clusters NGC 6791 and NGC 6819, 2) to extract their asteroseismic parameters, and 3) to determine their cluster membership.}{We followed a dedicated method based on difference imaging to extract the light curves of potential red giants in NGC 6791 and NGC 6819 from \emph{Kepler} superstamp data. We extracted the asteroseismic parameters of the stars that showed solar-like oscillations. We performed an asteroseismic membership study to identify which of these stars are likely to be cluster members.}{We found 149 red giant stars within the \emph{Kepler} superstamps, 93 of which are likely cluster members. We were able to find 29 red giants that are not primary targets of \emph{Kepler}, and therefore, their light curves had not been released previously. Five of these previously unknown red giants have a cluster membership probability greater than 95\%.}{}





   \maketitle
%
\section{Introduction}
Open clusters are remarkable features in the Milky Way. They contain up to a few thousand stars that are thought to have formed at the same time from the same molecular cloud. Hence, stars in clusters are assumed to have similar properties, such as age, initial chemical composition, and distance. Due to these shared initial conditions, studying the properties of cluster stars at different evolutionary stages allows us to constrain and test theories of stellar evolution and to gain insights into the formation and chemical evolution of our own Galaxy \citep[e.g.][]{vandenberg83,salaris13,cantat20}.
In this study, we search for additional oscillating red giants in the open clusters NGC 6791 and NGC 6819, which were observed by the \emph{Kepler} space mission \citep{kepler1,kepler2}. 

\emph{Kepler} observed the same field in the sky for four years. From the $\sim500\,000$ stars in the field that are brighter than the targeted threshold (<16 mag in the \emph{Kepler} passband; \citealt{total_stars}), $\sim$150\,000 stars were selected as primary targets for the mission. They are a subset of the \emph{Kepler} Input Catalog (KIC; \citealt{kic}). 
The \emph{Kepler} field of view contains four open clusters: NGC 6791, NGC 6811, NGC 6819, and NGC 6866.
Two of them, NGC 6791 and NGC 6819 ($\sim$8.5 and $\sim$2.5 Gyr old, respectively; \citealt{ages1}), were found to contain a substantial number of oscillating red giant stars, and $\sim$120 have been targeted by the \emph{Kepler} mission and were studied with asteroseismology based on these data. For these stars, studies have determined asteroseismic parameters, stellar properties such as mass and radii \citep{saskia11}, evolutionary stage \citep{Corsaro12}, mass loss \citep{Miglio12}, helium content \citep{mckeever19}, core mixing \citep{bossini17}, stellar ages \citep{ages1}, and cluster membership \citep{stello11}.
However, these clusters are likely to contain more oscillating stars that have not yet been analyzed. 

Asteroseismology is a powerful technique that can probe the internal structure of stars through the study of their global oscillations. This method can be used to obtain global properties of red giant stars, such as stellar mass, radius, and evolutionary stage. 
To perform an asteroseismic analysis, it is necessary to obtain the light curve of the star.
The mode of operation of \emph{Kepler} was based on providing the light curves for only those stars that were observed as primary targets of the mission, that is, a subset of the KIC catalog. 
In addition to the target pixel files that were obtained for the preselected stars, the mission collected photometric data of the whole central regions of the open clusters NGC 6791 and NGC 6819. These data products are so-called superstamps that comprise long-cadence ($\sim 30$~minutes) photometric observations with a spatial size of $200\times 200$ pixels (13.3 arcminutes on one side). These data can be used to extract the light curve of any resolved star within the superstamps. Most of the stars in the open clusters NGC 6791 and NGC 6819 were not primarily targeted, and therefore, no light curves were released. We aim to extract the light curves of oscillating red giants from the superstamp data that were not primary targets of \emph{Kepler}. 

The light curve of a star can be extracted with different methods. The most commonly used technique is aperture photometry, which was successfully applied to the observed KIC stars and some superstamp stars \citep[e.g.][]{kasoc,k2p2}. Aperture photometry is a reliable method for fairly isolated stars, but has its limits in highly crowded fields, for instance, central regions of open clusters or for saturated objects because it can introduce background signal.
In contrast, difference imaging is a preferred technique for probing crowded fields because it removes all nonvariable signal that neighboring stars might be introducing.
A light-curve extraction method based on difference imaging was implemented by \citet{iris}. They successfully tested their \emph{IRIS} pipeline on KIC targets that are located within the \emph{Kepler} superstamp data. 

We used the method developed by \citet{iris} to extract the light curves of red giant stars from the \emph{Kepler} open cluster superstamp data of NGC 6791 and NGC 6819. We selected $\sim 500$ stars in the superstamps that are potential cluster red giants. We did this using colors and magnitudes from {\it Gaia} data (eDR3; \citealt{edr3}), and distances derived by \citet{edr3_distances}. About half of the selected stars were not primary targets of \emph{Kepler}, and their light curves were not previously extracted. 
We downloaded the uncorrected light curves of the KIC targets in our sample, extracted by \citet{iris}, and applied the \emph{IRIS} pipeline to the non-KIC targets. 
We corrected the light curves using a modified version of the KASOC correction filter \citep{kasoc}. 
We performed an asteroseismic analysis including cluster membership study of those stars that showed solar-like oscillations.


\section{Target selection}
\label{sec:target_selection}
We focused on the open clusters NGC 6791 and NGC 6819, which are known to host a substantial number of red giant stars. The \emph{Kepler} mission provided  high-quality photometric data for these clusters over a time span of four years. These superstamp data enable an asteroseismic analysis of any resolved star within the central regions of the clusters, including stars that were not primary targets of the mission.

First, we computed the expected number of red giant stars in these clusters that are observable with \emph{Kepler}. Based on color-magnitude data from ground-based cluster surveys \citep{stetson,hole,ak}, we selected the stars that are likely to be red giants according to their location in the color-magnitude diagram. We only considered stars with a \emph{V} magnitude below 16.7, as we do not expect to find oscillations in fainter red giants \citep{stello11}.
We found $\sim200$ and $\sim150$ observable red giants in NGC 6791 and NGC 6819, respectively. So far, asteroseismic studies have focused on $\sim120$ oscillating red giants in total in these clusters, potentially leaving several dozen stars to be studied. 

To find these cluster red giant stars, we selected all the stars in the superstamps  that are within the published color, magnitude, and distance ranges of the confirmed cluster red giants \citep{stello11}. We obtained the colors and magnitudes for these reference stars from \emph{Gaia} eDR3. We used the \emph{gaia-kepler.fun} crossmatch database\footnote{\url{https://gaia-kepler.fun/}} to determine their \emph{Gaia} identification numbers from the KIC numbers provided by \citet{stello11}.
We took distance measurements provided by \citet{edr3_distances} that were derived based on \emph{Gaia} eDR3 parallaxes. 
The intervals of color, magnitude, and distance that we obtained for cluster red giants are shown in table \ref{table:1}.

\begin{table}
\caption{Intervals applied to our target selection.}             
\label{table:1}   
\centering                    
\begin{tabular}{c c c c}   
\hline\hline                
Cluster & Color $(bp-rp)$ & \emph{Gaia} (mag) & Distance (pc) \\    
\hline                       
   NGC 6791 & 1 -- 2.5 & 12 -- 17 & $2\,500$ -- $5\,500$ \\     
   NGC 6819 & 0.8 -- 2.5 & 10 -- 15    & $1\,000$ -- $4\,200$ \\
\hline                             
\end{tabular}
\tablefoot{Color, magnitude, and distance ranges for the confirmed cluster red giants in \citet{stello11}.}
\end{table}

\begin{figure*}
\centering
\includegraphics[width=0.9\hsize]{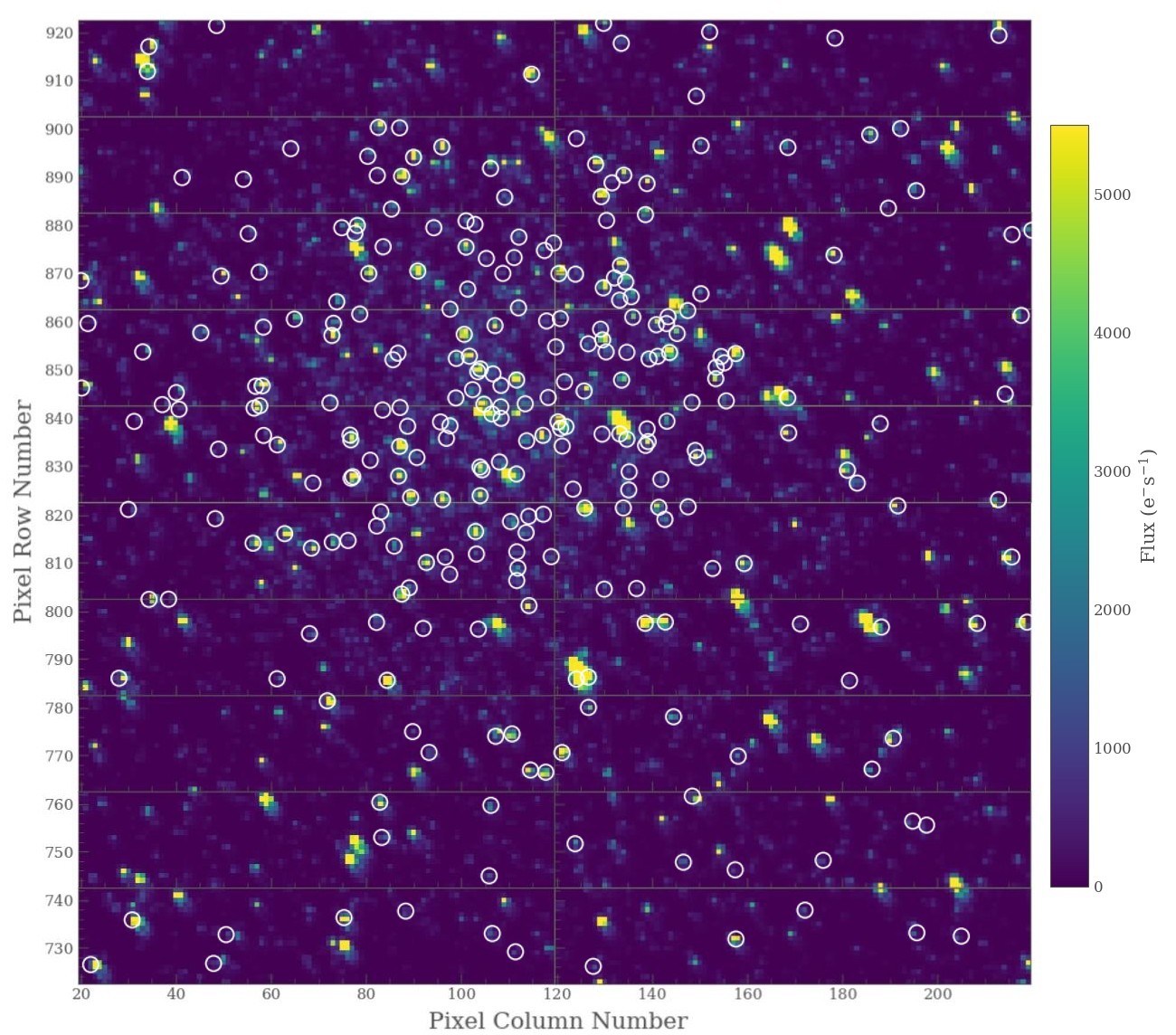}
\caption{Cluster superstamp of NGC 6791. These data were released as 20 substamps. We selected 281 stars as red giant candidates (encircled).}
\label{Superstamp of NGC 6791. }
\end{figure*}

\begin{figure*}
\centering
\includegraphics[width=0.9\hsize]{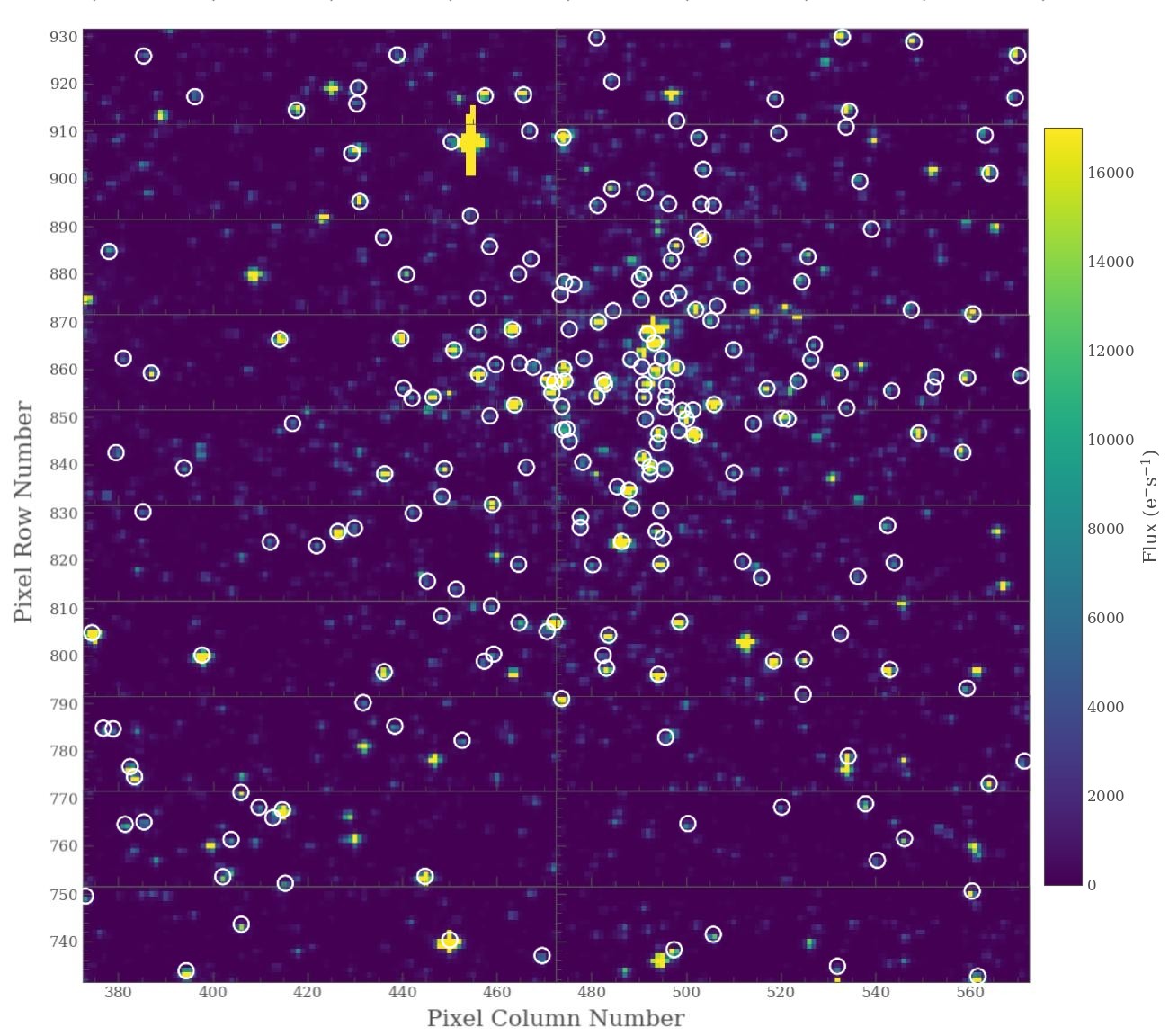}
\caption{Same as Fig.~\ref{Superstamp of NGC 6791. }, now for NGC 6819. We selected 232 stars as red giant candidates (encircled).}
\label{superstamp of NGC 6819}
\end{figure*}

We searched the \emph{Gaia} eDR3 catalog for stars that fulfilled these criteria within the superstamp regions. We obtained a list of 281 candidate red giants in NGC 6791 and 232 in NGC 6819. These are more than the $\sim200$ and $\sim150$ observable red giants that we expect in each cluster. Therefore, it is likely that a subset of these candidates are not red giant stars. Figures \ref{Superstamp of NGC 6791. } and \ref{superstamp of NGC 6819} show the positions of these candidates within the superstamps. Out of these stars, 225 are not primary targets of \emph{Kepler}, and their light curves have not been released so far. 


\section{Light-curve extraction}
\label{sec:data}
The \emph{Kepler} superstamps contain long-cadence ($\sim30$ minutes; \citealt{longcadence}) photometric data for  stars in NGC 6791 and NGC 6819, obtained during the observing quarters 1-17 of the mission. This results in over four years of data for both clusters. No observations are available for NGC 6819 during quarters 6, 10, and 14, due to a broken CCD.

To extract the light curves of the selected stars from the superstamps, we applied the \emph{IRIS} pipeline \citep{iris}, which is based on a difference-imaging method. The pipeline starts by cutting out a 5x5 pixel postage stamp from the superstamp with the target star at the center. This places the data in a format similar to a \emph{Kepler} target pixel file, that is, a time series of pixel images that contains all the photometric data of the star that were obtained during a given quarter. The code then estimates the crowding of the region around the star by comparing the total magnitude of the frame-averaged postage stamp to the known \emph{Kepler} magnitude of the star. If the two magnitudes are similar, the star is fairly isolated. However, if the postage magnitude is lower than the magnitude of the star (i.e., the postage-stamp flux is higher than the flux of the star), it is likely that a nearby star contaminates the data. Because \emph{Kepler} magnitudes are only available for primary targets of the mission \citep{kic}, we instead estimated the crowding using Gaia magnitudes because the two missions used similar passbands. Then, the 5×5 pixel postage stamp is resampled by a factor of 20 to 100×100 pixels using bilinear interpolation in order to create more precise apertures for the stars (see Fig. 3). 
Next, an aperture mask is defined as a 2D Gaussian function with a width based on the crowding of the region. Apertures for isolated stars have a larger width than those for stars with crowded backgrounds. This reduces the contamination of nearby stars in crowded regions.
Difference imaging is then applied to the 100×100 pixel stamps by subtracting the frame-averaged image from each frame in the stamps. This removes all the flux sources that are not variable. 
Finally, the flux is extracted from each frame by applying the aperture mask. The result is a light curve for a single quarter that only accounts for variable sources.

\begin{figure*}
    \centering
    \includegraphics[scale=0.3]{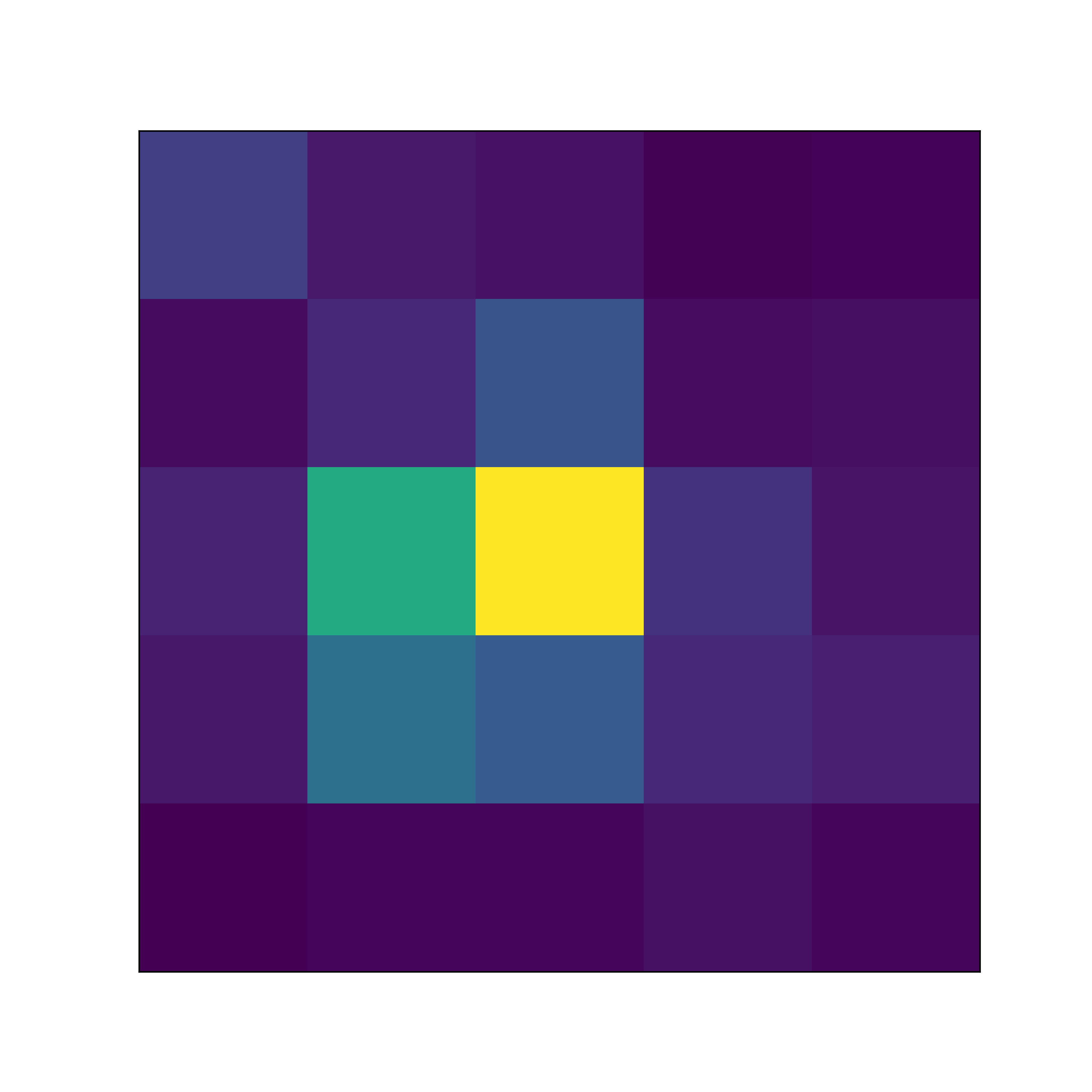}
    \includegraphics[scale=0.3]{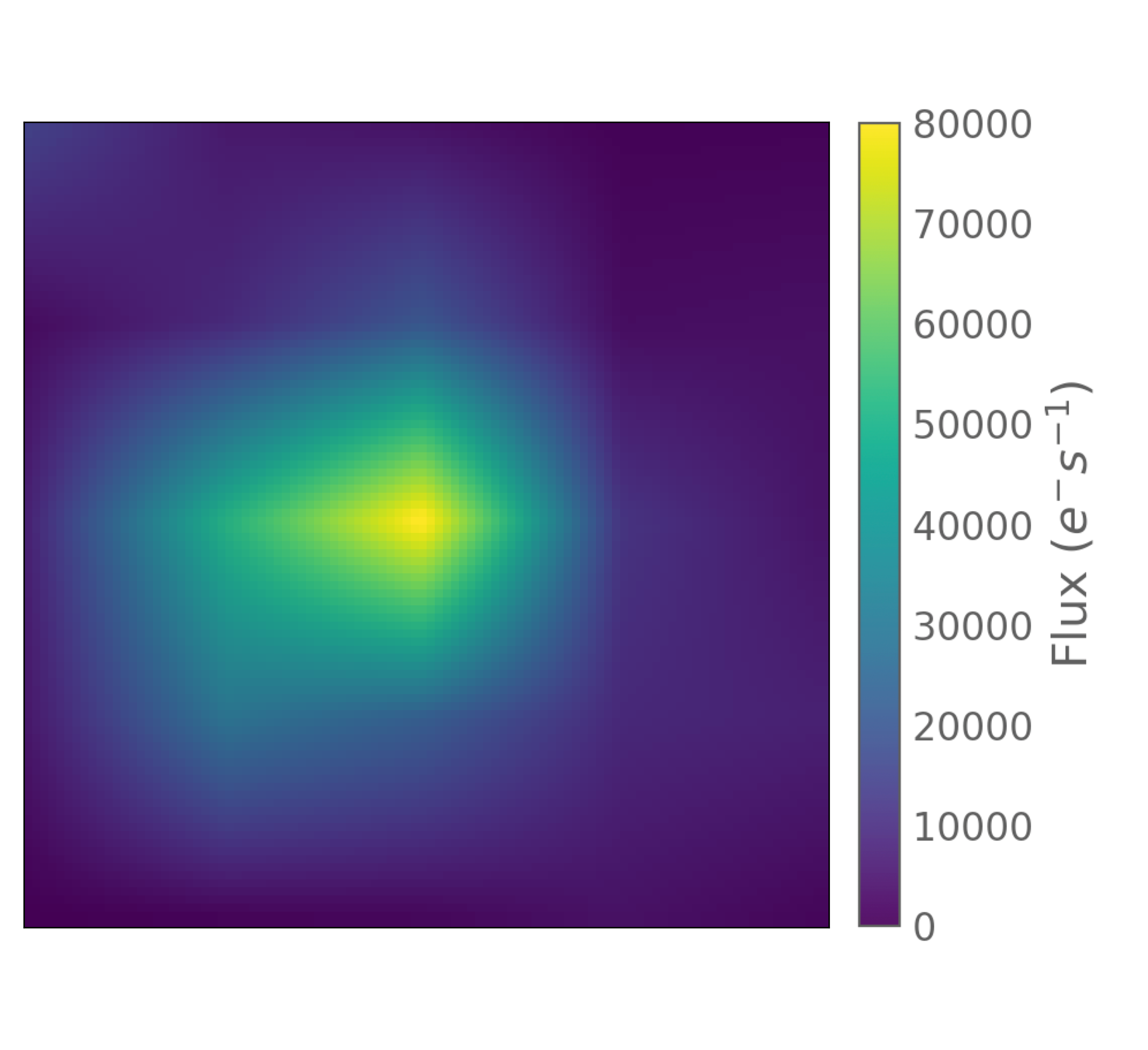}
    \caption{Resampling of the pixel stamps with the \emph{IRIS} code. Left: $5\times 5$ pixel stamp of one of the target stars in NGC 6819. Right: $100\times 100$ pixel stamp of the same star, obtained after resampling the original stamp by a factor of 20 using bilinear interpolation.}
    \label{FigInc}
     \end{figure*}

The light curves of the KIC targets in our sample have already been extracted using the \textit{IRIS} method by \citet{iris}. Thus, we downloaded the uncorrected light curves of the KIC stars and used the \textit{IRIS} pipeline to extract those of the non-KIC targets.
After extracting each light curve, we applied a number of corrections to remove systematics introduced by the instrument and the mission. We applied the {\sc KASOC} filter \citep{kasoc}, which was designed to correct \emph{Kepler} light curves obtained with aperture photometry. This pipeline uses the quality flags released by the mission and removes data points that are flagged as desaturation events, manual exclusions, and argabrightening. Furthermore, it corrects for the jumps in the light curve that are produced by attitude tweaks. Other quality flags are ignored. However, we realized that safe modes and Earth points produced significant jumps in our light curves. Therefore, we extended the jump corrections to data points showing these quality flags. Then, the pipeline filtered the light curve for long-term trends (i.e., instrumental
drifts and stellar activity) and short-term trends (i.e., transit features and instrumental effects, e.g., sudden jumps after safe modes), giving each filter a different weight depending on the relative standard deviation of the signal. Last, we performed sigma clipping to remove all data points that deviated by more than $5\sigma$ from the average.
The resulting light curves showed significantly diminished systematic effects.

After we obtained the corrected light curves of the star for every observed quarter, we normalized the flux to be shown in parts per million. Last, we appended the normalized light curves of each quarter to produce the final light curve.

The resulting light curves of the red giants in our catalog that show oscillations are available on
GitHub\footnote{\url{https://github.com/astroalba/kepler-red-giants-lc}}.

\section{Asteroseismic analysis}
\label{sec:astero}
We computed a power density spectrum for each star from the extracted light curves with the aim of detecting solar-like oscillations. The power density spectrum of a red giant star shows a distinct power excess that comprises the oscillation modes. 
To identify this potential power excess, we first implemented the automated CV method following \citet{Bell2019}. This method is based on the coefficients of variation (CV), that is, the ratio of the standard deviation of the power to its mean. This ratio should be equal to 1 in the presence of $\chi^{2}_{2}$ noise statistics as expected for the power density spectrum of solar-like oscillators. Any signal that is not caused by stochastic noise will present higher CV values. An expected width of the oscillatory signal in the CV method as a function of the frequency of maximum oscillation power ($\nu_{\rm max}$)  was introduced based on the analysis of the stars in the  APOKASC-DR2 \citep{pinsonneault18}.
We found 59 solar-like oscillators and 44 stars with ambiguous results. After visual inspection of the latter, we confirmed 46 oscillations that the CV method could not clearly detect, probably due to changes in the statistics of the power density spectrum due to contamination of neighboring stars. Combining this with oscillating red giants from the literature and a final visual inspection, we identified a final sample of 149 oscillating red giants, (91 in NGC\,6791 and 58 in NGC\,6819) in our set of 513 stars.

\begin{figure}
\centering
\includegraphics[width=\hsize]{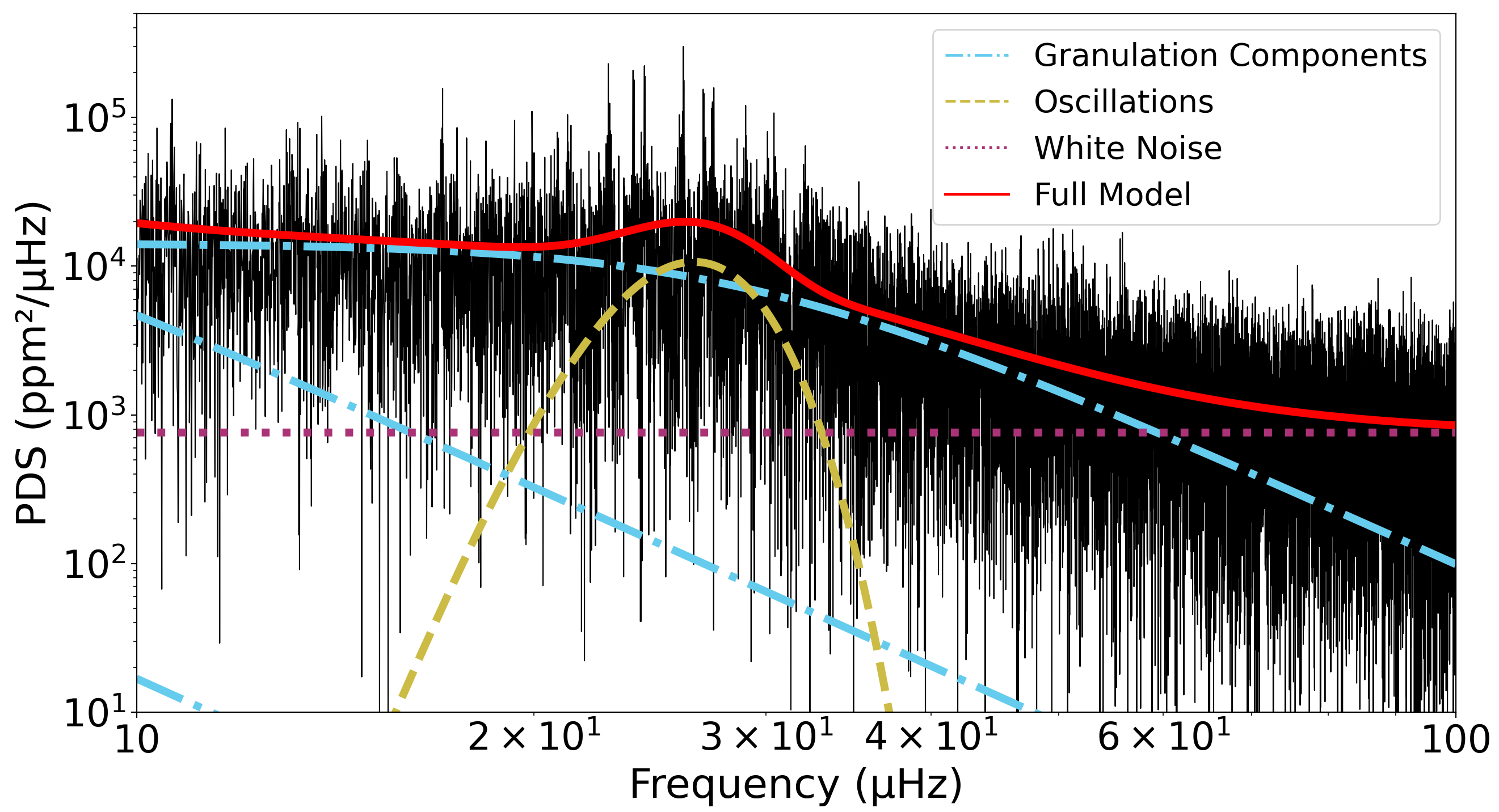}
\caption{Power density spectrum (in black) of one of the target stars in NGC 6791 (star 39 in table \ref{table:1}). The solid red line shows the global model fit computed by TACO, which includes three granulation components (dashed blue lines), one white-noise component (horizontal dotted magenta line), and a Gaussian fit to the oscillation power excess (dashed yellow line).}
\label{pds}
\end{figure}

For these 149 oscillating stars, we performed an asteroseismic analysis and extracted the frequency of their maximum oscillation  power ($\nu_\textrm{max}$) using the code called tools for the automated characterization of oscillations (TACO) (Hekker et al. in prep.).
We obtained $\nu_\textrm{max}$ from a global fit to the power density spectrum through a Bayesian Markov chain Monte Carlo framework. The model consists of three granulation components, one white-noise term, and a Gaussian fit to the power excess comprising $\nu_\textrm{max}$. We took the 50th percentile from the posterior $\nu_\textrm{max}$ distribution as the 
final parameter estimate and the 16th and 84th percentiles as its uncertainties.
In Figure~\ref{pds} we show a power density spectrum with the global fit.
We report the \emph{Gaia} eDR3 $G$ magnitudes and the measured $\nu_\textrm{max}$ values for the oscillating red giant stars in the field of view of NGC 6791 and NGC 6819 in tables \ref{6791} and \ref{6819}.

\begin{figure}
\centering
\includegraphics[width=\hsize]{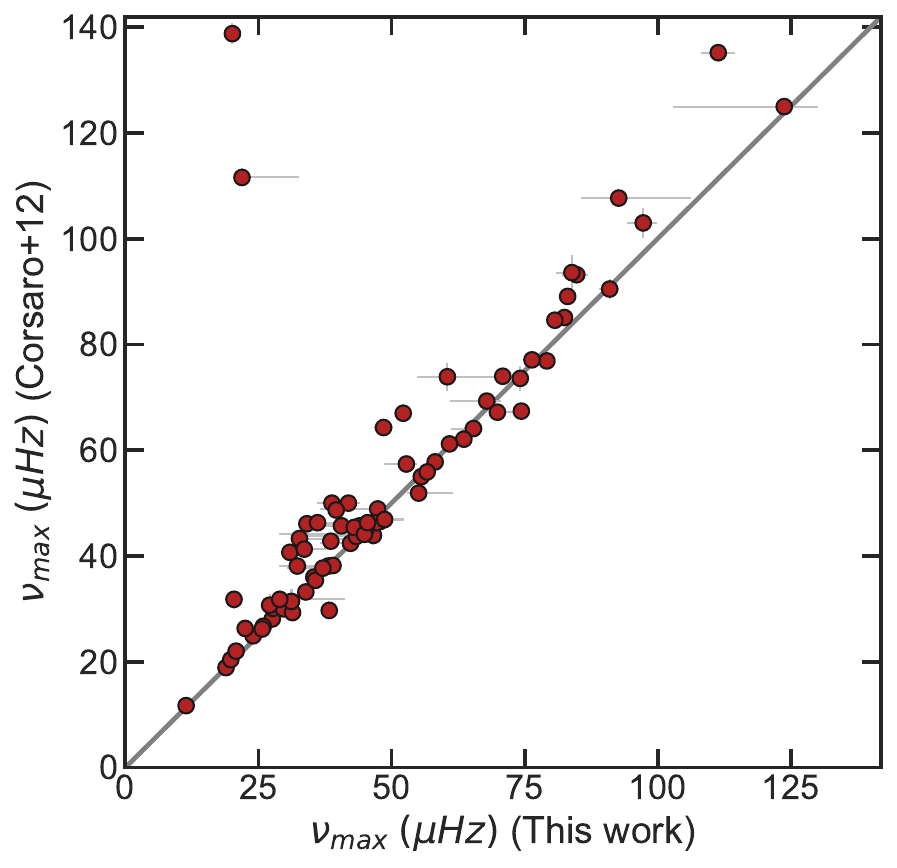}
\caption{Comparisons of the frequencies of maximum oscillation ($\nu_\textrm{max}$) that we obtained to those released by \citet{Corsaro12}.}
\label{corsaro}
\end{figure}

Out of the 149 oscillating red giants, 83 were previously analyzed by \citet{Corsaro12}. They obtained asteroseismic parameters, including $\nu_\textrm{max}$, for 115 oscillating red giant stars in the \emph{Kepler} open clusters. At the time, only seven quarters of \emph{Kepler} data were available, which corresponds to a time stamp of $\sim19$ months (less than half of the data that are currently available). A comparison between our results and theirs is shown in figure \ref{corsaro}. Although the two results agree for most stars (with a sigma-clipped scatter of 4, in line with earlier works; e.g. \citealt{comparisons_hekker}), there are a few stars with very different values of $\nu_\textrm{max}$ as a result of the two methods. These outliers seem to have picked up an entirely different signal on the power density spectrum. This might be due to differences in the methods used for light-curve extraction and correction.

\section{Cluster membership study}
\label{sec:memberships}
We performed an asteroseismic cluster membership study to identify the oscillating stars that belong to NGC 6791 and NGC 6819.
Following the approach introduced by \citet{stello10}, we used the $\nu_{\rm max}$ measurements together with the {\it Gaia} $G$ magnitudes to determine the most likely cluster members. We note here that the $G$ passband covers a wavelength range of about 330 -- 1050\,nm, which is similar to the 2MASS passbands used by \citet{stello10}.

For red giant stars, the asteroseismic parameter $\nu_\textrm{max}$ is related to the global properties of the star through a scaling relation \citep{eq1,eq2},
\begin{equation}
    \nu_\textrm{max}\simeq\frac{M/ {\rm M}_\odot(T_\textrm{eff}/{T_\textrm{eff,}}_\odot)^{3.5}}{L/\rm L_\odot}\nu_\textrm{max,\sun},
\end{equation}
where $M$, $T_\textrm{eff}$, and $L$ are the mass, effective temperature, and luminosity of the star. The nominal effective temperature of the Sun is $T_{\textrm{eff,}\odot}=5\,772\,K$ \citep{teff}, and $\nu_\textrm{max,\sun}=3\,100\,\mu \textrm{Hz}$ \citep{numaxsun} is a reference value that scales the relation to the Sun.

For red giant stars, the stellar masses and effective temperatures vary only marginally compared to the luminosities, which have a wider range of values when these stars evolve. Thus, there is a tight correlation between $\nu_\textrm{max}$ and the stellar luminosity. For cluster red giants, this translates into a correlation between $\nu_\textrm{max}$ and the apparent stellar magnitude because all the stars are assumed to be located at a similar distance. By plotting the apparent magnitude versus $\nu_\textrm{max}$, cluster red giants line up, while fore- and background stars 
deviate from the correlation.

\begin{figure}
\centering
\includegraphics[width=\hsize]{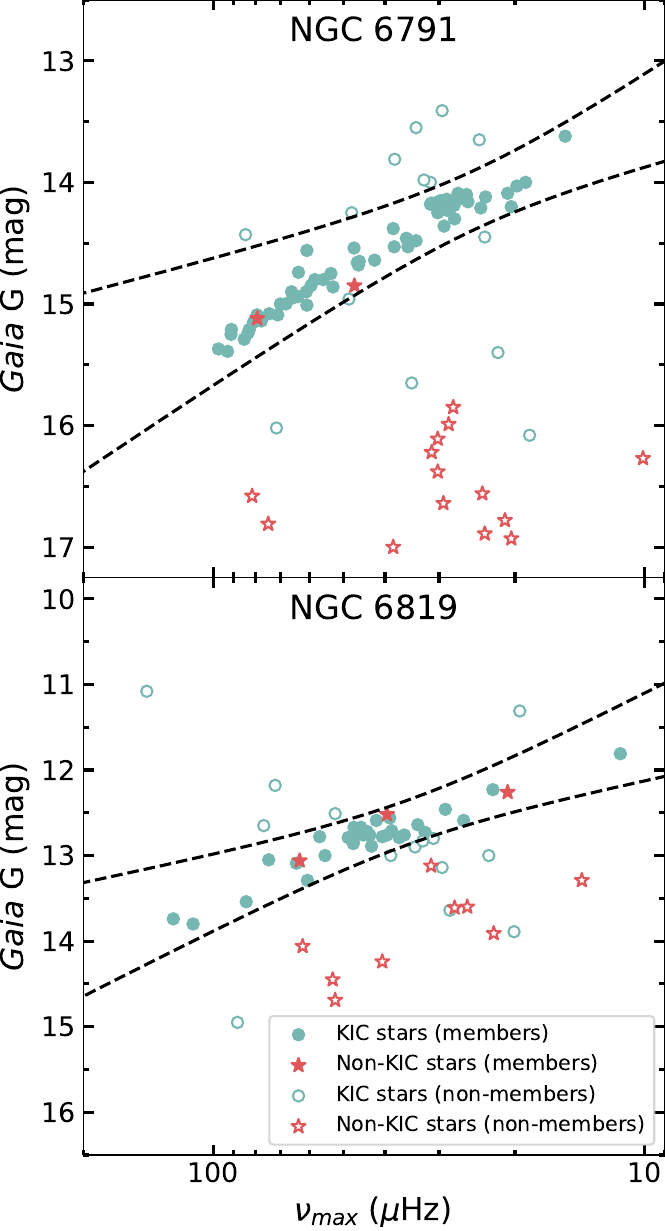}
\caption{\emph{Gaia} G magnitude vs. frequency of the maximum oscillation power $\nu_{\rm max}$ for red giants in NGC 6791 and NGC 6819. Dots are KIC stars, and star symbols are newly detected stars. Filled symbols are likely to be cluster members, and open symbols are nonmembers. The axes are oriented to make the plot similar to a color-magnitude diagram. The dashed lines show the regions in which cluster stars tend to line up in these diagrams and their 95\% confidence intervals.}
\label{member1}
\end{figure}

In Figure \ref{member1} we show the measured asteroseismic $\nu_\textrm{max}$ values versus the published {\it Gaia} $G$ magnitudes for our stars with detected solar-like oscillations in NGC 6791 (top) and NGC 6819 (bottom). In each diagram, we indicate the confidence region (dashed lines) hosting stars with a 95\% confidence to be part of the cluster population. The confidence intervals for each cluster were computed around a linear regression line based on $\nu_\textrm{max}$ values, $G$ magnitudes, and their respective uncertainties for a set of preselected KIC open cluster red giants that were studied by \citet{stello10}. Their cluster memberships were independently confirmed through asteroseismic, radial velocity, and proper motion studies. The cluster membership probabilities of stars outside the 95\% confidence intervals decrease with increasing distance to the confidence region.

\begin{figure}
\centering
\includegraphics[width=\hsize]{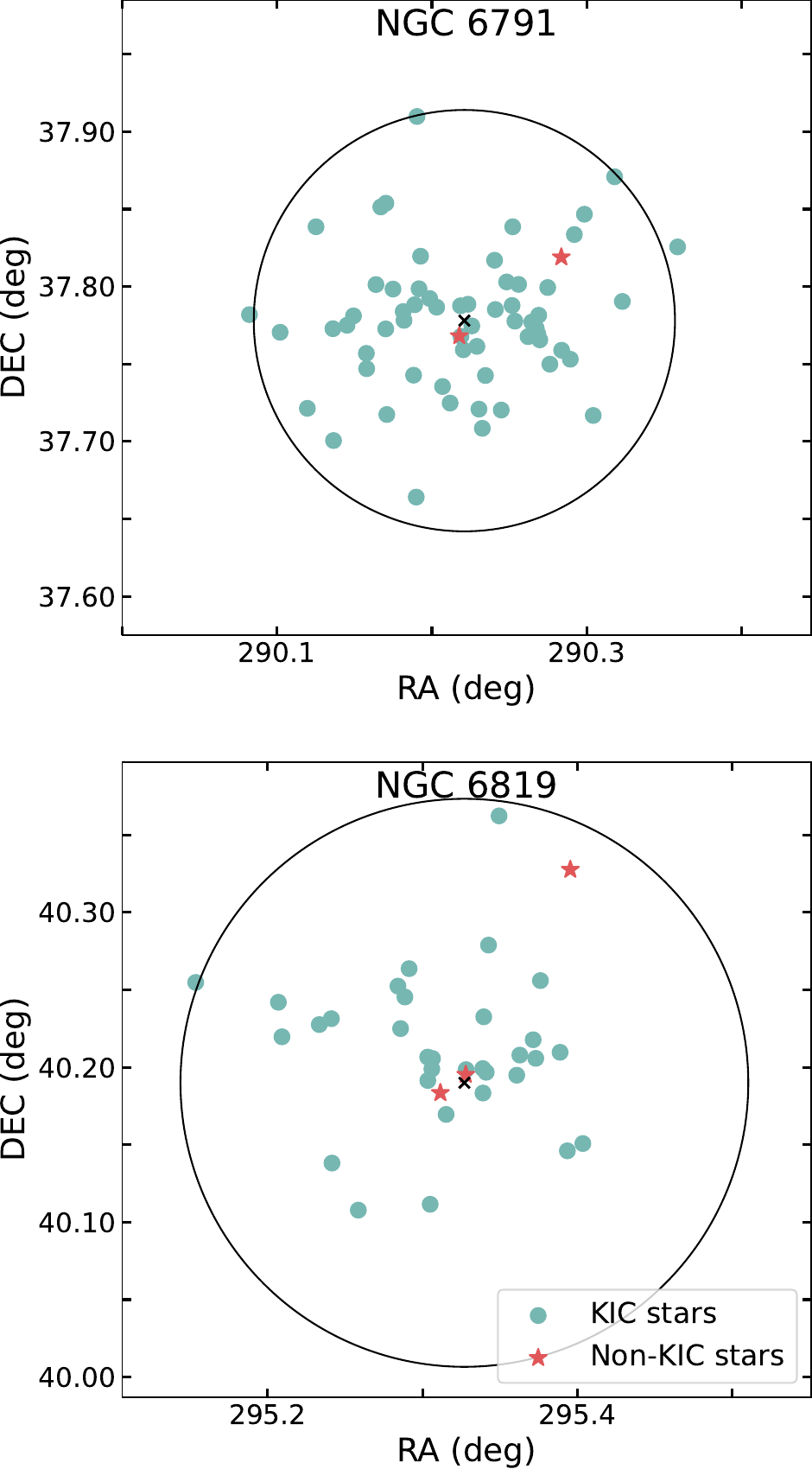}
\caption{Sky positions of the oscillating red giants that we found to be likely cluster members. Dots are KIC stars, and star symbols are newly detected stars. The large circles represent the approximate location of the clusters \citep{cantat20}, and the central cross represents the cluster center.}
\label{fig:positions}
\end{figure}

Using this method, we found that 93 red giants have an asteroseismic cluster membership probability of more than 95\% (61 in NGC 6791 and 32 in NGC 6819). Of these, 5 stars are not primary targets of \emph{Kepler}, and they are confirmed to be cluster members using an asteroseismic analysis for the first time here. Figure \ref{fig:positions} shows the positions of these oscillating cluster red giants within their respective clusters. 

Because we expect these clusters to host $\sim200$ and $\sim150$ observable red giant stars, respectively (see section \ref{sec:target_selection}), it is likely that we missed about three-quarters of the cluster red giants. This could be due to our restrictive selection in color, magnitude, and distance (see table \ref{table:1}), or due to limitations of our method for extracting the oscillations in the most crowded regions of the cluster, where the aperture masks are narrow in order to prevent contamination. It is also possible that we missed some oscillating stars due to a high level of noise in the power density spectrum, which would partially hide the oscillations.

An alternative method for determining cluster membership that does not rely on asteroseismology was presented by \citet{cantat20}. This method applies an unsupervised classification scheme using \textit{Gaia} DR2 proper motions and parallaxes to compute membership probabilities. The authors studied $1\,481$ stellar clusters in the Galactic disk, including NGC 6791 and NGC 6819, and provided a list of cluster stars and their membership probability. We added the membership determination from \citet{cantat20} to tables \ref{6791} and \ref{6819} of our oscillating stars. Out of the 149 oscillating red giants that we detected, 113 are considered cluster members with a probability of 100\% in their work, while our method retrieved 93 members with a probability of 95\%. Thus, the two methods agree for most of the stars in our sample, except for $\sim10\%$ of them. It seems that our method is much more restrictive in determining cluster membership, and we might be missing some members due to this.
\section{Summary and conclusions}
\label{sec:conclusions}
We have successfully extracted the light curves of 513 red giant candidates in the open clusters NGC 6791 and NGC 6819 using the superstamp data from \emph{Kepler}. We were able to confirm that 149 of these stars are oscillating red giants, and we measured their frequency of maximum oscillation power ($\nu_\textrm{max}$). By performing an asteroseismic membership study, we found that 93 of these red giants are likely to be cluster members. 75 of these were previously reported to be cluster members by \citet{stello11} based on a similar membership determination method. 11 nonmember red giants in our sample were reported to be cluster members by \citet{stello11}. This is probably because our membership determination method is based on the correlation between the \emph{Gaia} magnitude of each star and its $\nu_\textrm{max}$, while they used the \emph{K} magnitude against $\Delta\nu$. Although both strategies are based on the same principle, it is possible that there is some scatter between the two methods, which would explain the disagreement for the 11 stars. It is also worth noting that at the time, only four quarters of \textit{Kepler} data were available, making their light curves four times shorter than those we extracted.

Out of the 149 oscillating red giants that we found, 29 are not primary targets of \emph{Kepler}, and their light curves have not been released previously. We found that 5 of these red giants are asteroseismic cluster members. 

The method that we followed to extract the light curves from the \emph{Kepler} superstamps allowed us to obtain the asteroseismic parameters of stars that are very close to the cluster center, which would be very challenging for aperture photometry. This method cannot only be applied to solar-like oscillators, but also to any other star within the superstamp data. Moreover, it can be used to extract the light curves in any moderately crowded region as long as high-precision long-term photometric data are available. 

In the upcoming PLATO mission, the provisional fields that are expected to be observed (LOPS2 and LOPN1) contain about 300 open clusters \citep{platoclusters}. Our method can be used to extract the light curves of cluster stars observed by PLATO and provides an opportunity of performing asteroseismic analyses on these clusters. These analyses would allow us to obtain the parameters of more clusters in the Milky Way, which in turn would bring us closer to understanding the nature of these objects in our Galaxy. 

\begin{acknowledgements}
We acknowledge funding from the ERC Consolidator Grant DipolarSound (grant agreement \# 101000296). This work was funded by the Deutsche Forschungsgemeinschaft (DFG, German Research Foundation) -- Project-ID 138713538 -- SFB 881 (``The Milky Way System'', subproject P02)
\end{acknowledgements}

\bibliographystyle{aa}
\bibliography{references}

\begin{appendix}
\onecolumn
\section{Tables}
 \begin{longtable}{ccccccccc}
 \caption{\label{6791} Oscillating red giants in NGC 6791.}\\
 \hline\hline
 ID& KIC & RA (\textdegree)& DEC (\textdegree)& \emph{Gaia} (mag) & $\nu_\mathrm{max}$ ($\mu\mathrm{Hz}$) & Member (this work) & S11 member & C20 member\\
 \hline
 \endfirsthead
 \caption{continued.}\\
 \hline\hline
 ID& KIC & RA (\textdegree)& DEC (\textdegree)& \emph{Gaia} (mag) & $\nu_\mathrm{max}$ ($\mu\mathrm{Hz}$) & Member (this work) & S11 member& C20 member\\
 \hline
 \endhead
 \hline
 \endfoot
1          & -       & 290.1856 & 37.6687   & 16.93 & $20.4_{-0.6}^{+0.4}$  &No & -&- \\
2          & -       & 290.3508 & 37.7717   & 16.27 & $10.1_{-0.3}^{+0.6}$  &No & -&Yes\\
3          & -       & 290.2810 & 37.7486   & 16.89 & $23.5_{-1.9}^{+10.2}$ &No & - &Yes\\
4          & -       & 290.2403 & 37.7394   & 16.38 & $30.2_{-0.5}^{+0.4}$ &No & - &-\\
5          & -       & 290.2641 & 37.7722   & 16.06 & $30.2_{-1.0}^{+8.7}$  &No & - &Yes\\
6          & -       & 290.2548 & 37.7754   & 16.56 & $23.8_{-0.2}^{+0.4}$  &No & - &Yes\\
7          & -       & 290.2796 & 37.7958   & 16.22 & $31.2_{-0.7}^{+0.5}$ &No & - &Yes\\
8          & -       & 290.2794 & 37.7974   & 16.78 & $21.1_{-0.04}^{+0.1}$  &No & - &Yes\\
9          & -       & 290.1708 & 37.7726   & 15.85 & $27.8_{-11.0}^{+1.7}$  &No & - &Yes\\
10         & -       & 290.0669 & 37.8447   & 17.00 & $38.3_{-1.4}^{+1.4}$  &No & - &-\\
11         & -       & 290.2177 & 37.7681   & 14.85 & $47.1_{-1.5}^{+1.3}$ &Yes & - &-\\
12         & -       & 290.2525 & 37.7804   & 16.64 & $29.3_{-1.3}^{+1.2}$ &No & - &Yes\\
13         & -       & 290.2031 & 37.7875   & 15.99 & $28.5_{-4.7}^{+0.6}$ &No & - &Yes\\
14         & -       & 290.2500 & 37.8004   & 16.81 & $74.6_{-1.9}^{+1.4}$  &No & - &Yes\\
15         & -       & 290.2836 & 37.8190   & 15.12 & $79.1_{-2.5}^{+1.0}$  &Yes & - &Yes\\
16         & -       & 290.2318 & 37.8126   & 16.58 & $81.3_{-13.1}^{+0.8}$ &No & - &Yes\\
17         & -       & 290.1668 & 37.8011   & 16.08 & $18.5_{-2.1}^{+0.8}$  &No & - &-\\
18         & 2297384 & 290.1899 & 37.6641   & 14.21 & $20.4_{-0.3}^{+0.3}$  &Yes &Yes &Yes\\
19         & 2435987 & 290.0822 & 37.7819   & 14.53 & $38.1_{-0.4}^{+0.4}$  &Yes &Yes &Yes\\
20         & 2436097 & 290.1020 & 37.7705   & 14.66 & $42.3_{-5.8}^{+0.6}$  &Yes &Yes &Yes\\
21         & 2436209 & 290.1196 & 37.7214   & 14.80 & $58.2_{-1.0}^{+0.7}$  &Yes &Yes &Yes\\
22         & 2436332 & 290.1361 & 37.7728   & 14.30 & $27.6_{-0.5}^{+0.7}$   &Yes &Yes &Yes\\
23         & 2436334 & 290.1365 & 37.7006   & 14.78 & $63.5_{-7.3}^{+3.1}$  &Yes & - &Yes\\
24         & 2436417 & 290.1452 & 37.7750   & 14.10 & $25.9_{-0.4}^{+0.5}$  &Yes &Yes &Yes\\
25         & 2436458 & 290.1494 & 37.7811   & 14.53 & $35.4_{-0.3}^{+0.7}$  &Yes &Yes &Yes\\
26         & 2436540 & 290.1577 & 37.7569   & 14.86 & $52.8_{-4.3}^{+2.0}$  &Yes &Yes &Yes\\
27         & 2436543 & 290.1579 & 37.7470   & 14.12 & $23.4_{-0.7}^{+0.7}$  &Yes & - &Yes\\
28         & 2436593 & 290.1618 & 37.7430   & 15.40 & $21.9_{-1.1}^{+10.8}$  &No &Yes &Yes\\
29         & 2436608 & 290.1634 & 37.7437   & 14.45 & $23.5_{-0.4}^{+0.4}$  &No & - &Yes\\
30         & 2436682 & 290.1702 & 37.7727   & 14.14 & $27.1_{-0.5}^{+0.5}$   &Yes & - &Yes\\
31         & 2436688 & 290.1709 & 37.7174   & 15.09 & $79.2_{-1.0}^{+1.1}$  &Yes &Yes &Yes\\
32         & 2436732 & 290.1749 & 37.7984   & 14.14 & $28.8_{-1.8}^{+0.5}$  &Yes &Yes &Yes\\
33         & 2436814 & 290.1815 & 37.7839   & 14.21 & $24.0_{-0.4}^{+0.8}$  &Yes &Yes &Yes\\
34         & 2436824 & 290.1820 & 37.7782   & 14.48 & $33.9_{-0.3}^{+0.3}$  &Yes &Yes &Yes\\
35         & 2436900 & 290.1882 & 37.7428   & 14.46 & $35.7_{-0.4}^{+0.3}$  &Yes &Yes &Yes\\
36         & 2436912 & 290.1889 & 37.7883   & 14.18 & $30.0_{-1.0}^{+1.0}$  &Yes &Yes &Yes\\
37         & 2436935 & 290.1905 & 37.7406   & 15.65 & $34.7_{-3.9}^{+1.1}$ &No & - &Yes\\
38         & 2436944 & 290.1917 & 37.7986   & 14.17 & $28.5_{-0.5}^{+0.5}$ &Yes &Yes &-\\
39         & 2437040 & 290.1986 & 37.7923   & 14.16 & $25.7_{-0.3}^{+0.3}$ &Yes &Yes &Yes\\
40         & 2437103 & 290.2031 & 37.7867   & 14.38 & $38.3_{-1.4}^{+1.4}$ &Yes &Yes &Yes\\
41         & 2437164 & 290.2069 & 37.7355   & 14.16 & $30.1_{-0.7}^{+0.5}$ &Yes & - &Yes\\
42         & 2437240 & 290.2118 & 37.7248   & 14.65 & $45.9_{-1.0}^{+0.5}$   &Yes &Yes &Yes\\
43         & 2437267 & 290.2141 & 37.7751   & 13.81 & $38.0_{-1.5}^{+2.0}$ &No & - &Yes\\
44         & 2437296 & 290.2159 & 37.7800   & 16.02 & $71.4_{-0.7}^{+0.7}$  &No & - &Yes\\
45         & 2437325 & 290.2185 & 37.7876   & 15.29 & $84.8_{-2.7}^{+2.1}$   &Yes &Yes &Yes\\
46         & 2437327 & 290.2188 & 37.7679   & 14.85 & $59.4_{-0.9}^{+1.0}$    &Yes & - &Yes\\
47         & 2437353 & 290.2203 & 37.7592   & 14.14 & $28.8_{-0.5}^{+0.4}$  &Yes &Yes &Yes\\
48         & 2437402 & 290.2233 & 37.7886   & 14.66 & $46.4_{-0.6}^{+0.6}$  &Yes &Yes &Yes\\
49         & 2437444 & 290.2258 & 37.7746   & 14.00 & $18.9_{-0.3}^{+0.2}$ &Yes &Yes &Yes\\
50         & 2437488 & 290.2291 & 37.7614   & 14.95 & $65.4_{-4.2}^{+1.3}$  &Yes &Yes &Yes\\
51         & 2437507 & 290.2304 & 37.7209   & 14.03 & $19.8_{-0.3}^{+0.4}$  &Yes &Yes &Yes\\
52         & 2437539 & 290.2326 & 37.7085   & 14.54 & $47.2_{-12.7}^{+0.8}$  &Yes & - &Yes\\
53         & 2437564 & 290.2346 & 37.7426   & 14.23 & $28.07_{-0.6}^{+0.7}$  &Yes &Yes &Yes\\
54         & 2437589 & 290.2363 & 37.7187   & 14.25 & $47.8_{-0.6}^{+0.6}$ &No &Yes &Yes\\
55         & 2437653 & 290.2410 & 37.7852   & 15.09 & $70.9_{-1.0}^{+1.0}$ &Yes &Yes &Yes\\
56         & 2437698 & 290.2448 & 37.7203   & 14.19 & $27.7_{-0.7}^{+1.1}$ &Yes &Yes &Yes\\
57         & 2437781 & 290.2518 & 37.7878   & 15.21 & $82.5_{-1.4}^{+1.6}$ &Yes &Yes &Yes\\
58         & 2437805 & 290.2536 & 37.7777   & 14.15 & $29.7_{-0.6}^{+0.6}$  &Yes &Yes &Yes\\
59         & 2437901 & 290.2621 & 37.7677   & 14.25 & $30.2_{-0.4}^{+0.4}$  &Yes & - &Yes\\
60         & 2437933 & 290.2645 & 37.7771   & 15.39 & $92.7_{-7.0}^{+13.5}$ &Yes &Yes &Yes\\
61         & 2437957 & 290.2674 & 37.7730   & 15.25 & $91.0_{-2.0}^{+1.8}$  &Yes &Yes &Yes\\
62         & 2437972 & 290.2684 & 37.7694   & 15.15 & $80.7_{-0.9}^{+0.8}$ &Yes &Yes &Yes\\
63         & 2437976 & 290.2690 & 37.7815   & 15.24 & $83.1_{-1.7}^{+1.5}$  &Yes &Yes &Yes\\
64         & 2437987 & 290.2697 & 37.7656   & 14.18 & $31.4_{-0.6}^{+0.6}$  &Yes &Yes &Yes\\
65         & 2438038 & 290.2748 & 37.7994   & 14.94 & $63.6_{-0.6}^{+0.6}$  &Yes  &Yes &Yes\\
66         & 2438051 & 290.2762 & 37.7499   & 14.09 & $27.1_{-0.6}^{+0.5}$  &Yes &Yes &Yes\\
67         & 2438069 & 290.2777 & 37.7176   & 14.00 & $31.4_{-1.0}^{+1.1}$ &No & - &-\\
68         & 2438139 & 290.2836 & 37.7971   & 13.55 & $33.9_{-2.2}^{+1.0}$ &No & - &No (70\%)\\
69         & 2438140 & 290.2838 & 37.7589   & 15.00 & $67.9_{-6.9}^{+2.7}$ &Yes &Yes &Yes\\
70         & 2438191 & 290.2894 & 37.7531   & 15.21 & $90.8_{-1.9}^{+1.9}$&Yes & - &Yes\\
71         & 2438333 & 290.3042 & 37.7168   & 14.90 & $60.9_{-1.2}^{+0.9}$ &Yes &Yes &Yes\\
72         & 2438470 & 290.3230 & 37.7904   & 14.90 & $65.9_{-1.7}^{+0.6}$ &Yes & - &Yes\\
73         & 2568575 & 290.0295 & 37.8334   & 13.41 & $29.5_{-1.0}^{+3.6}$ &No & - &-\\
74         & 2569055 & 290.1252 & 37.8386   & 14.17 & $31.2_{-0.6}^{+1.1}$ &Yes &Yes &Yes\\
75         & 2569078 & 290.1288 & 37.8125   & 13.65 & $24.2_{-3.6}^{+0.3}$ &No & - &-\\
76         & 2569360 & 290.1639 & 37.8013   & 14.09 & $20.8_{-0.3}^{+0.3}$ &Yes &Yes &Yes\\
77         & 2569390 & 290.1670 & 37.8514   & 15.01 & $60.7_{-1.2}^{+1.1}$ &Yes & - &Yes\\
78         & 2569426 & 290.1703 & 37.8538   & 14.56 & $60.7_{-0.9}^{+2.7}$&Yes & - &-\\
79         & 2569618 & 290.1927 & 37.8196   & 14.80 & $55.6_{-0.5}^{+0.5}$ &Yes &Yes &Yes\\
80         & 2570094 & 290.2405 & 37.8170   & 15.00 & $69.9_{-1.4}^{+4.2}$ &Yes &Yes &Yes\\
81         & 2570172 & 290.2483 & 37.8030   & 15.08 & $74.2_{-1.2}^{+0.7}$ &Yes &Yes &Yes\\
82         & 2570214 & 290.2522 & 37.8386   & 14.11 & $26.1_{-0.9}^{+1.1}$ &Yes &Yes &Yes\\
83         & 2570244 & 290.2560 & 37.8014   & 15.37 & $97.3_{-3.1}^{+2.6}$ &Yes &Yes &Yes\\
84         & 2570384 & 290.2752 & 37.8681   & 14.96 & $48.5_{-1.7}^{+1.8}$ &No &Yes &Yes\\
85         & 2570518 & 290.2920 & 37.8336   & 14.68 & $46.1_{-8.4}^{+0.6}$ &Yes &Yes &Yes\\
86         & 2570575 & 290.2985 & 37.8467   & 15.14 & $77.4_{-1.0}^{+1.0}$ &Yes & - &Yes\\
87         & 2570578 & 290.2990 & 37.8794   & 13.98 & $32.5_{-1.8}^{+0.6}$ &No & - &-\\
88         & 2570652 & 290.3113 & 37.8854   & 14.43 & $84.2_{-0.8}^{+0.8}$ &No & - &No (60\%)\\
89         & 2570696 & 290.3180 & 37.8709   & 13.62 & $15.3_{-0.2}^{+0.2}$ &Yes & - &-\\
90         & 2570924 & 290.3587 & 37.8256   & 14.75 & $53.4_{-0.5}^{+0.5}$ &Yes & - &Yes\\
91         & 2707716 & 290.1904 & 37.9098   & 14.36 & $29.2_{-1.3}^{+1.1}$  &Yes & - &Yes\\
\end{longtable}
\tablefoot{We report whether they are cluster members following our method (this work), and whether they were determined to be cluster members in \citealt{stello11} (S11) and \citealt{cantat20} (C20; only when their membership probability is 100\%).}

 \begin{longtable}{ccccccccc}
 \caption{\label{6819} Oscillating red giants in NGC 6819.}\\
 \hline\hline
 ID& KIC & RA (\textdegree)& DEC (\textdegree)& \emph{Gaia} (mag) & $\nu_\mathrm{max}$ ($\mu\mathrm{Hz}$) & Member (this work) & S11 member & C20 member\\
 \hline
 \endfirsthead
 \caption{continued.}\\
 \hline\hline
 ID& KIC & RA (\textdegree)& DEC (\textdegree)& \emph{Gaia} (mag) & $\nu_\mathrm{max}$ ($\mu\mathrm{Hz}$) & Member (this work) & S11 member& C20 member\\
 \hline
 \endhead
 \hline
 \endfoot
 92         & -       & 295.2984 & 40.0901   & 13.60 & $25.8_{-1.2}^{+1.0}$  & No &-&-\\
93         & -       & 295.2936 & 40.1458   & 14.06 & $62.1_{-1.6}^{+1.5}$  & No  &-&-\\
94         & -       & 295.2583 & 40.1445   & 14.24 & $40.6_{-0.9}^{+0.9}$  & No  &-&-\\
95         & -       & 295.2676 & 40.1656   & 13.61 & $27.6_{-0.6}^{+1.1}$  & No &-&-\\
96         & -       & 295.3279 & 40.1953   & 12.26 & $20.8_{-0.5}^{+0.4}$  & Yes  &-&No (60\%)\\
97         & -       & 295.2844 & 40.1772   & 13.29 & $14.0_{-0.2}^{+0.2}$  & No  &-&-\\
98         & -       & 295.3115 & 40.1835   & 12.52 & $39.6_{-0.4}^{+0.7}$  & Yes  &-&Yes\\
99         & -       & 295.4381 & 40.1639   & 14.69 & $52.2_{-0.4}^{+0.4}$  & No  &-&-\\
100         & -       & 295.4468 & 40.2657   & 14.45 & $52.9_{-13.4}^{+2.3}$  & No &- &-\\
101        & -       & 295.3954 & 40.3276   & 13.06 & $63.1_{-1.1}^{+1.0}$  & Yes &- &-\\
102        & -       & 295.3926 & 40.3291   & 13.12 & $31.3_{-0.9}^{+15.0}$ & No  &-&-\\
103        & -       & 295.3621 & 40.3200   & 13.91 & $22.4_{-0.03}^{+0.2}$  & No &-&-\\
104        & 4936825 & 295.2691 & 40.0808   & 13.14 & $29.5_{-0.8}^{+2.9}$  & No & -&-\\
105        & 5023953 & 295.2415 & 40.1382   & 12.59 & $41.9_{-2.1}^{+1.8}$  & Yes &Yes &Yes\\
106        & 5024043 & 295.2585 & 40.1078   & 12.78 & $56.7_{-1.9}^{+1.6}$  & Yes &Yes&-\\
107        & 5024268 & 295.2959 & 40.1866   & 11.31 & $19.5_{-1.5}^{+6.0}$  & No &No&Yes\\
108        & 5024297 & 295.3010 & 40.1927   & 12.90 & $34.1_{-1.0}^{+13.02}$ & No &Yes&Yes\\
109        & 5024312 & 295.3034 & 40.1915   & 13.54 & $83.9_{-3.0}^{+2.7}$  & Yes &Yes&Yes\\
110        & 5024322 & 295.3049 & 40.1116   & 13.09 & $64.2_{-21.0}^{+1.4}$  & Yes &-&No (60\%)\\
111        & 5024327 & 295.3060 & 40.1989   & 12.75 & $46.6_{-0.6}^{+0.6}$  & Yes &Yes&Yes\\
112        & 5024404 & 295.3152 & 40.1696   & 12.86 & $47.4_{-10.9}^{+1.7}$  & Yes&Yes&Yes\\
113        & 5024414 & 295.3164 & 40.1865   & 12.65 & $76.4_{-0.2}^{+0.2}$  & No &Yes&Yes\\
114        & 5024456 & 295.3210 & 40.1811   & 11.08 & $142.6_{-0.4}^{+0.4}$  & No  &Yes&Yes\\
115        & 5024476 & 295.3240 & 40.1544   & 12.51 & $52.2_{-1.5}^{+1.6}$  & No &Yes&Yes\\
116        & 5024512 & 295.3281 & 40.1985   & 13.29 & $60.5_{-5.8}^{+13.8}$ & Yes&Yes&Yes\\
117        & 5024517 & 295.3288 & 40.1947   & 13.00 & $38.8_{-2.8}^{+5.3}$  & No &Yes&Yes\\
118        & 5024582 & 295.3389 & 40.1992   & 12.67 & $47.2_{-0.6}^{+0.6}$  & Yes&Yes&No (30\%)\\
119        & 5024583 & 295.3390 & 40.1834   & 12.73 & $32.3_{-3.5}^{+5.4}$  & Yes&Yes&Yes\\
120        & 5024601 & 295.3411 & 40.1968   & 12.46 & $29.0_{-3.2}^{+12.3}$ & Yes&Yes&Yes\\
121        & 5024750 & 295.3608 & 40.1949   & 11.81 & $11.4_{-0.2}^{+0.4}$  & Yes &Yes&Yes\\
122        & 5024967 & 295.3935 & 40.1461   & 12.79 & $48.7_{-3.0}^{+3.6}$  & Yes &Yes&Yes\\
123        & 5025021 & 295.4035 & 40.1508   & 12.84 & $47.4_{-0.6}^{+0.5}$  & Yes &-&-\\
124        & 5025101 & 295.4205 & 40.1867   & 13.00 & $23.0_{-0.5}^{+0.2}$  & No &-&-\\
125        & 5111718 & 295.1535 & 40.2548   & 13.80 & $111.4_{-3.2}^{+3.1}$  & Yes&Yes&Yes\\
126        & 5111767 & 295.1655 & 40.2312   & 13.64 & $28.3_{-0.9}^{+2.2}$  & No  &-&-\\
127        & 5111940 & 295.2069 & 40.2420   & 13.00 & $55.1_{-2.1}^{+6.4}$  & Yes &Yes&Yes\\
128        & 5111949 & 295.2092 & 40.2197   & 12.76 & $39.6_{-1.7}^{+1.5}$  & Yes&Yes&Yes\\
129        & 5111987 & 295.2166 & 40.2600   & 14.95 & $87.9_{-11.8}^{+2.1}$  & No &- &-\\
130        & 5112072 & 295.2333 & 40.2276   & 13.74 & $123.8_{-20.9}^{+6.3}$  & Yes &Yes&Yes\\
131        & 5112118 & 295.2411 & 40.2314   & 12.59 & $26.3_{-1.3}^{+0.9}$  & Yes &-&-\\
132        & 5112281 & 295.2695 & 40.2476   & 12.18 & $71.9_{-0.6}^{+0.6}$  & No  &-&-\\
133        & 5112361 & 295.2841 & 40.2524   & 13.05 & $74.4_{-1.1}^{+1.1}$  & Yes &No&-\\
134        & 5112373 & 295.2858 & 40.2250   & 12.76 & $43.4_{-12.1}^{+ 0.8}$  & Yes&Yes&Yes\\
135        & 5112387 & 295.2886 & 40.2454   & 12.78 & $40.6_{_5.1}^{+1.3}$  & Yes &Yes&Yes\\
136        & 5112401 & 295.2913 & 40.2637   & 12.56 & $39.0_{-0.4}^{+0.4}$  & Yes &Yes&Yes\\
137        & 5112403 & 295.2919 & 40.2190   & 13.89 & $20.1_{-1.0}^{+1.1}$  & No  &Yes&Yes\\
138        & 5112467 & 295.3033 & 40.2066   & 12.76 & $36.1_{-0.8}^{+8.1}$  & Yes &Yes&Yes\\
139        & 5112491 & 295.3064 & 40.2057   & 12.71 & $45.2_{-0.5}^{+0.5}$  & Yes &Yes&Yes\\
140        & 5112730 & 295.3395 & 40.2326   & 12.72 & $44.0_{-7.3}^{+1.3}$  & Yes&Yes&Yes\\
141        & 5112734 & 295.3407 & 40.2031   & 12.80 & $30.9_{-1.2}^{+1.4}$  & No &Yes&Yes\\
142        & 5112744 & 295.3426 & 40.2789   & 12.89 & $43.0_{-0.6}^{+0.5}$  & Yes &Yes &Yes\\
143        & 5112880 & 295.3628 & 40.2079   & 12.23 & $22.5_{-1.5}^{+1.8}$  & Yes&Yes &Yes\\
144        & 5112938 & 295.3714 & 40.2178   & 12.76 & $44.9_{-16.0}^{+2.3}$  & Yes &Yes&Yes\\
145        & 5112948 & 295.3726 & 40.2389   & 12.83 & $32.7_{-0.6}^{+9.9}$  & No &Yes&Yes\\
146        & 5112950 & 295.3731 & 40.2058   & 12.71 & $38.6_{-1.2}^{+1.0}$  & Yes &Yes&Yes\\
147        & 5112974 & 295.3761 & 40.2560   & 12.64 & $33.6_{-2.4}^{+4.1}$  & Yes &Yes&Yes\\
148        & 5113041 & 295.3888 & 40.2097   & 12.79 & $37.1_{-6.7}^{+1.1}$  & Yes&Yes&Yes\\
149        & 5200152 & 295.3494 & 40.3623   & 12.67 & $45.5_{-1.8}^{+1.0}$  & Yes&Yes&Yes\\
\end{longtable}
\tablefoot{We report whether they are cluster members following our method (this work), and whether they were determined to be cluster members in \citealt{stello11} (S11) and \citealt{cantat20} (C20; only when their membership probability is 100\%).}

\end{appendix}
\end{document}